\documentclass[sigconf,natbib=false,screen,balance]{acmart}
\usepackage{graphicx}  
\usepackage{comment}
\usepackage{subcaption}
\usepackage{xurl}
\title{TRACE: Topology-aware Reconstruction of Accidents \\ in CARLA for AV Evaluation }

\author{Nahian Salsabil} 
\email{abj6hv@virginia.edu}
\orcid{ }
\affiliation{%
  \institution{University of Virginia}
  \city{Charlottesville}
  \state{Virginia}
  \country{USA}
}

\author{Sebastian Elbaum}
\email{selbaum@virginia.edu}
\orcid{0000-0001-9592-1352}
\affiliation{%
  \institution{University of Virginia}
  \city{Charlottesville}
  \state{Virginia}
  \country{USA}
}

\newcounter{secounter}

\newcounter{nscounter}

\newcommand{\tool}{TRACE }

\copyrightyear{2026}
\acmYear{2026}
\setcopyright{cc}
\setcctype{by}
\acmConference[FSE Companion '26]{34th ACM Joint European Software Engineering Conference and Symposium on the Foundations of Software Engineering}{July 05--09, 2026}{Montreal, QC, Canada}
\acmBooktitle{34th ACM Joint European Software Engineering Conference and Symposium on the Foundations of Software Engineering (FSE Companion '26), July 05--09, 2026, Montreal, QC, Canada}
\acmDOI{10.1145/3803437.3806426}
\acmISBN{979-8-4007-2636-1/2026/07}

\begin{document}

\begin{abstract}
Validating Autonomous Vehicles (AVs) requires exposure to rare, safety-critical scenarios, infrequent in routine driving data. Existing benchmarks address this by generating synthetic conflicts or mapping accident descriptions to abstract road geometries, failing to capture the topological complexity of real-world crashes. We introduce \tool, a pipeline that automates the reconstruction of NHTSA crash reports into high-fidelity CARLA simulations by (1) retrieving site-specific OpenStreetMap data to preserve exact road topology, (2) leveraging Large Language Models to infer vehicles' initial state from road geometry and pre-crash maneuvers, and (3) generating simulation trajectories from semi-structured report data. Using this pipeline, we curated a benchmark of 52 diverse accident scenarios covering varied collision types, road topologies, and pre-crash maneuvers, providing a challenging open-source resource for testing AV systems against real-world failures.
\end{abstract}



\maketitle

\section{Introduction}

The successful and responsible deployment of Autonomous Vehicles (AVs) relies on a robust evaluation process. Benchmarks are fundamental to this effort: they standardize evaluation, enable comparisons, and reveal system weaknesses.
For AVs, this systematic evaluation is essential to ensure system dependability prior to widespread public adoption.

This necessity has led to numerous evaluation suites\cite{words2col, safebench, robuste2e, chatscene} and tools \cite{Gen_Adv, lctgen, surfalgan, AC3R, sovar, accidentsim} to create such benchmarks  utilizing high-fidelity simulators like CARLA\cite{carla_web}. While early benchmarks primarily assess routine driving competencies, recent efforts have proposed to include safety-critical scenarios generated from accident reports. However, the few open-source specialized frameworks available oversimplify accidents, rendering limited reproduction fidelity in terms of road topologies, vehicle trajectories, and collision types. This limits the strength of software engineering methods towards evaluation and validation.

To address these limitations in evaluation realism and critical scenario validation coverage, we have developed  \tool, a  pipeline that systematically derives simulation scenarios from the NHTSA Crash Data\cite{NHTSA_Crash_API}. This pipeline involves transformations of real-world maps to fit CARLA's format and of analytical and statistical analyses to reconstruct the conditions present in crash reports. More specifically, it estimates the precise initial states (position and orientation) of the involved vehicles, and generates the specific trajectories that lead to the collision described in a crash report.
In parallel, the pipeline retrieves the real map where the accident occurred. It then  automatically instantiates the map and spawns the vehicles involved in the crash to reproduce the crash within the CARLA simulator. 
Our pipeline seeks the generation of scenarios that exhibit three key properties: (1) retain the road topology and signage of the accident site
and (2) spawn the vehicles with the state and planned trajectories that lead to the reported accident sequence.

Using this pipeline, we have constructed and manually curated a benchmark of 52 scenarios that exhibit diverse  topology, vehicle trajectory, and crash type as defined by the FARS (Fatality Analysis
Reporting System) Manual \cite{NHTSA_FARS_2025}. Each scenario package includes the original accident report, a summary of the crash, its corresponding map as retrieved by our pipeline, and the generated CARLA simulation scenario, which includes vehicles' starting position, direction, speed, and trajectory waypoints. By providing this collection of contextually rich, accident-derived scenarios, this benchmark is intended to offer a more challenging and realistic resource for testing the robustness of autonomous driving systems. 
The TRACE benchmark and pipeline used in this paper are available at \url{https://doi.org/10.5281/zenodo.19431924}, with the latest updates at \url{https://github.com/NahianSalsabil/carla-benchmark}.

We envision \tool serving three use cases: 1) \textbf{Testing AVs collision avoidance.} where researchers can replace vehicle control systems with their own to assess collision avoidance capabilities. 2) \textbf{Stress testing perception and prediction.} Developers can focus on the identified crash-prone  topologies that have specific intersection geometries or visual occlusions to evaluate how they affect their agent's behavior or their ability to anticipate the erratic trajectories of other vehicles. 3) \textbf{Automated Test-Case Generation}. \tool facilitates the generation of diverse safety-critical training scenarios to address the scarcity of long-tail data required to train robust learned components.
 
\begin{figure}
    \centering
    \begin{subfigure}[b]{0.2\textwidth}
        \includegraphics[width=\textwidth]{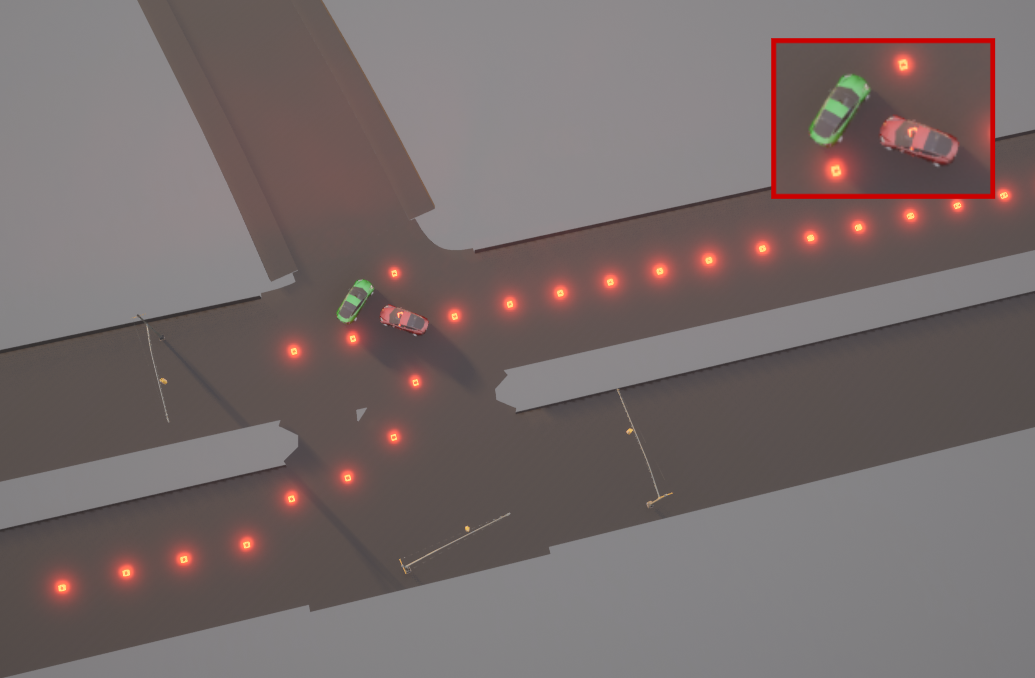}
        \caption{At a T-intersection.}
    \end{subfigure}
    \hfill
    \begin{subfigure}[b]{0.2\textwidth}
        \includegraphics[width=\textwidth]{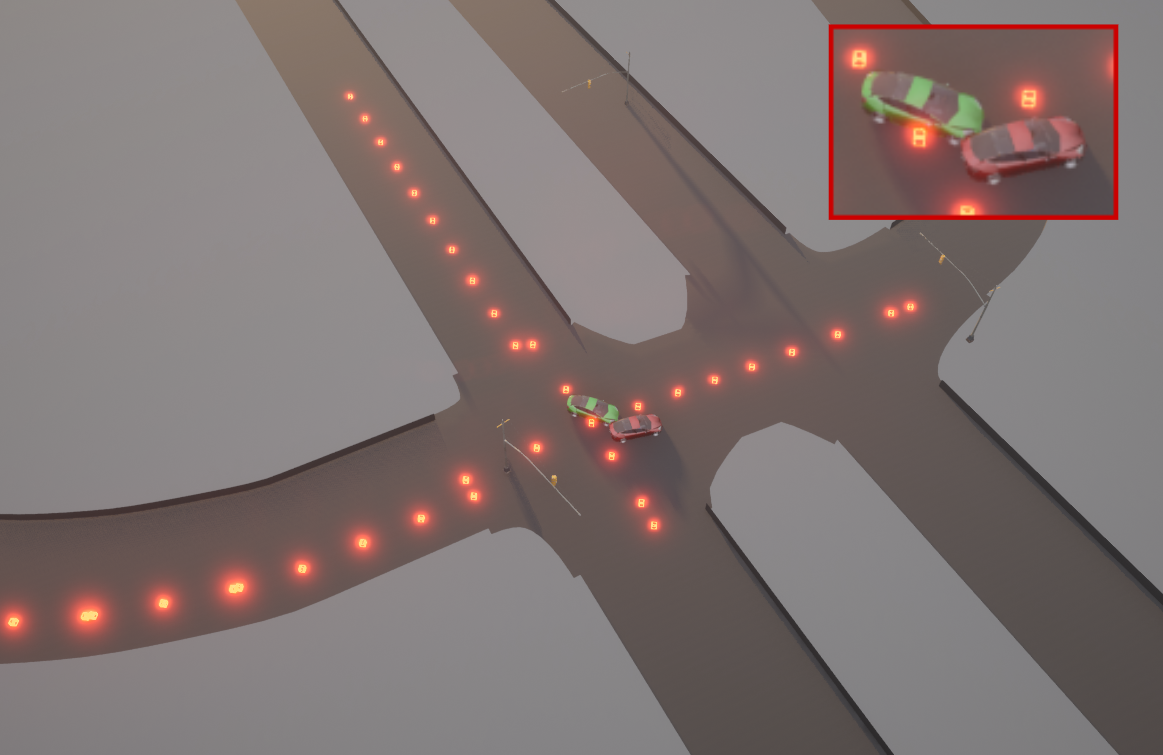}
        \caption{At a 4-way intersection.}
    \end{subfigure}
    \hfill
    \begin{subfigure}[b]{0.2\textwidth}
        \includegraphics[width=\textwidth]{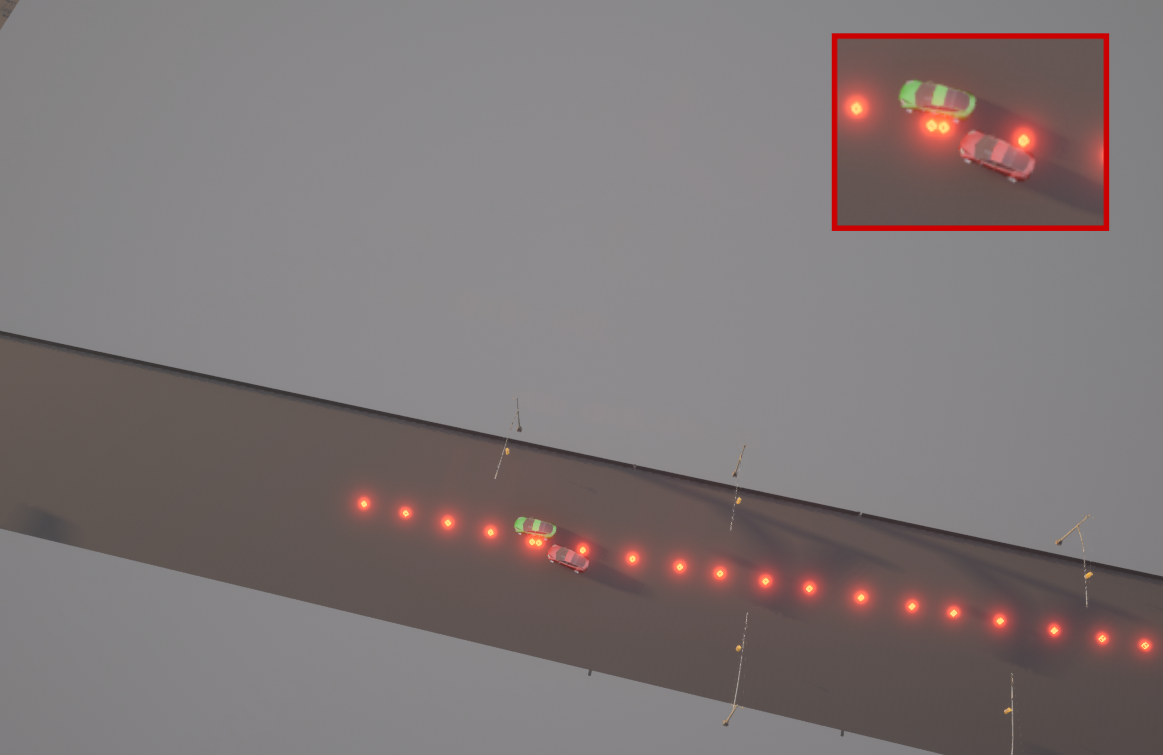}
        \caption{While a vehicle  stopped.}
    \end{subfigure}
    \hfill
    \begin{subfigure}[b]{0.2\textwidth}
        \includegraphics[width=\textwidth]{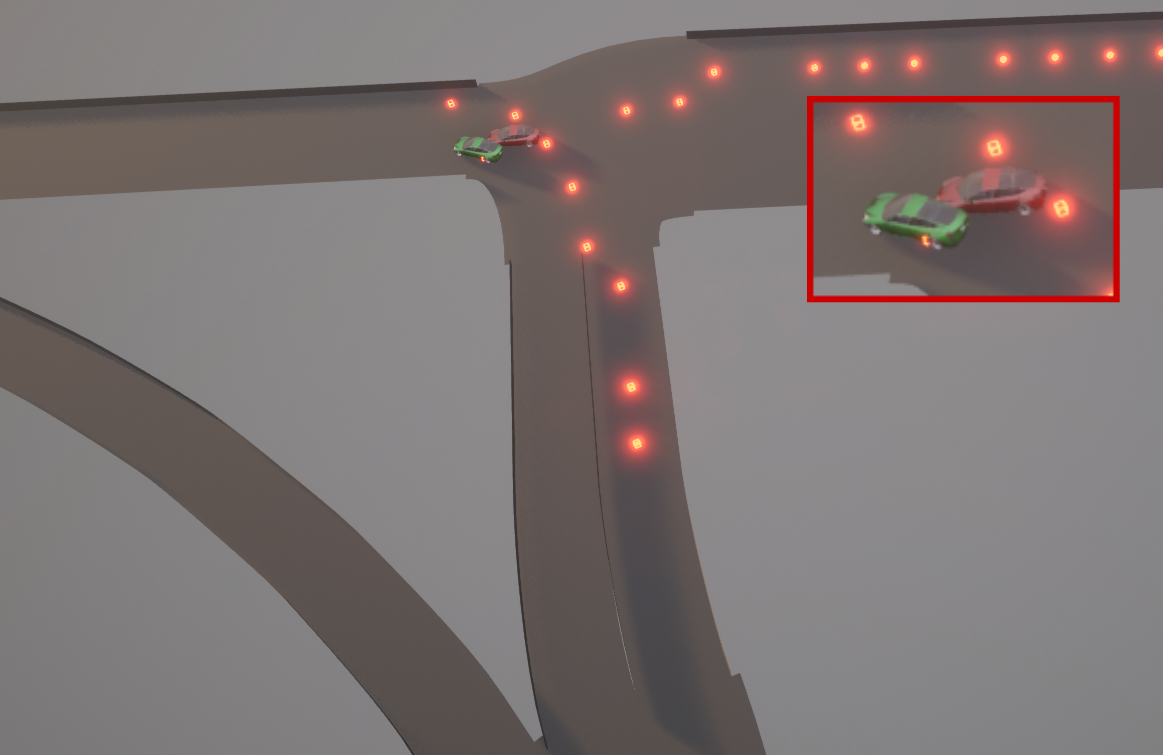}
        \caption{Changing traffic ways.}
    \end{subfigure}
    \hfill
    \begin{subfigure}[b]{0.2\textwidth}
        \includegraphics[width=\textwidth]{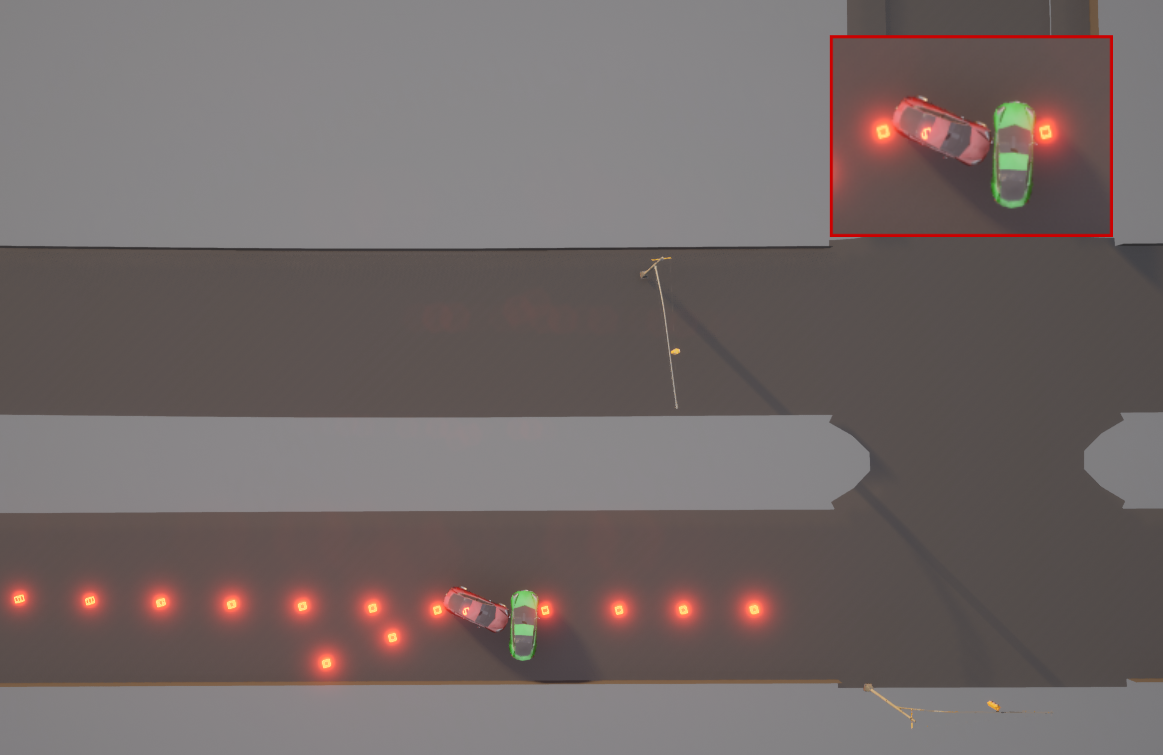}
        \caption{At an Angle.}
    \end{subfigure}
    \hfill
    \begin{subfigure}[b]{0.2\textwidth}
        \includegraphics[width=\textwidth]{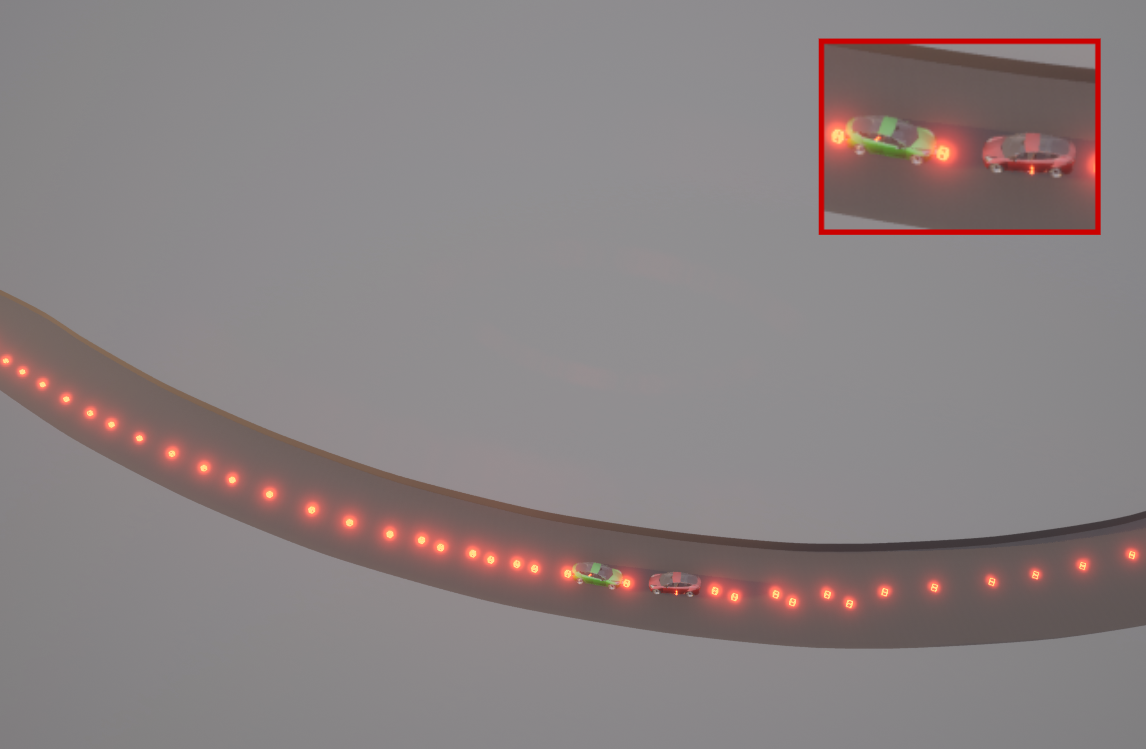}
        \caption{Curve in Front-to-Front.}
    \end{subfigure}
    \hfill
    \caption{Sample scenarios generated by \tool showing the diversity of  topologies,  collisions, and vehicle trajectories. } 
 \end{figure}
 
\section{Background}
\label{sec:background}

Our work relates to approaches generating safety-critical scenarios to test AVs and more specifically, to those aiming to increase the scenario realism by incorporating data from crash reports. 

Techniques for autonomous driving simulation are generally classified into three categories: 
(1) adversarial generation  \cite{Gen_Adv, Gen_Acc_Prone, closed_loop, stein23adv}, which creates challenging environments through malicious agent behaviors but often lacks computational efficiency and diversity; 
(2)  knowledge-based generation  \cite{ontology, CARLA_Leaderboard, regression}, which utilizes predefined rules and constraints yet struggles with implementation complexity and the inability to capture the full spectrum of high-risk, non-compliant behaviors; and 
(3)  data-driven generation  \cite{cmts, safebench, surfalgan, woodlief25diff}, which leverages real-world data but is frequently limited by the scarcity of safety-critical events in standard driving datasets.  Our work extends the data-driven approach by generating scenarios directly from real-world crash reports, thereby bypassing the ``long-tail" distribution problem to focus on confirmed collision events.

\begin{figure}[t]
    \centering
    \begin{minipage}{\linewidth}
        \centering
        \includegraphics[width=0.8\linewidth]{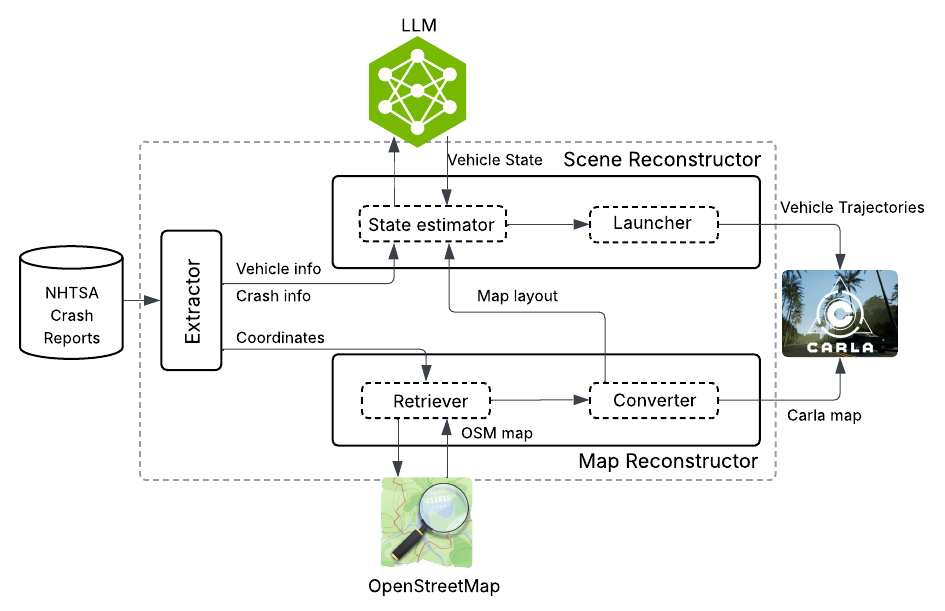}
    \end{minipage}
        \caption{\tool 
        pipeline. The \textit{Extractor}  parses NHTSA crash reports. The \textit{Map Reconstructor} retrieves and converts OSM data to OpenDRIVE. The \textit{Scene Reconstructor} uses an LLM-based State Estimator to infer vehicle trajectories for the \textit{Launcher} to execute in simulation.}
        \label{fig:pipeline}       
 \end{figure}

Scenario generation approaches that incorporate elements from crash reports can add realism. Numerous datasets of crash reports are available, such as the NHTSA Crash Investigation System\cite{NHTSA_CISS}, the California Autonomous Vehicle Collision Reports\cite{CA_DMV_AV_Reports}, 
and the European Road Safety Observatory\cite{erso}.

Existing techniques for generating scenarios from crash reports vary significantly in fidelity and accessibility. Early tools like AC3R \cite{AC3R} and SoVAR \cite{sovar} are publicly available but rely on abstract geometries or generic map-matching, which fail to capture precise real-world topologies; additionally, AC3R uses ad-hoc mechanism for data extraction and only focuses on the impact, while SoVAR’s template-based extraction limits its ability to handle diverse accident descriptions. More recent frameworks like CrashAgent \cite{crashagent} and AccidentSim \cite{accidentsim} utilize MLLMs and high-fidelity simulators like CARLA and LGSVL, yet their validation is often restricted to synthetic layouts, and neither their code nor datasets are available for verification. SAFE \cite{safe} is available and similar in its use of MLLMs, but it relies on a domain-specific language (DSL) to synthesize road environments, which again lacks the geospatial target fidelity. Our approach overcomes these limitations by (1) reconstructing exact road networks via OpenStreetMap, (2) employing LLM-supported state estimation and adaptive trajectory generation to handle diverse crash types, and (3) providing an open-source tool and benchmark to ensure reproducibility.

\section{\tool : Crash Reconstruction Pipeline}
\label{sec:pipeline}
 
Our first contribution   is a pipeline that  automates the reproduction of real crash scenarios in the CARLA simulator. As illustrated in \autoref{fig:pipeline}, the process starts with an \textit{Extractor} module that consumes NHTSA crash reports. The \textit{Extractor} then produces two outputs leading to two parallel processing streams, one for \textit{Scenario Reconstruction} and one for \textit{Map Reconstruction}. These streams are later integrated as a map with the vehicles involved in the crash spawned in CARLA. 

\subsection{Crash Data Extraction}
The pipeline begins with the \textbf{Extractor} module, which ingests raw crash reports from the NHTSA\cite{NHTSA_Crash_API} database. While the initial data originates from Police Accident Reports (PARs) filed by  law enforcement officers, these records are subsequently standardized  by state-level FARS analysts  to check for accuracy and consistency. They are extensive, semi-structured XML documents containing from 1000 to 1500 data fields spanning accident, vehicle, driver, occupant, damage, and fatality details, some of which are subjective and often incomplete. 
The Extractor parses these reports to isolate three distinct categories of information:
1) \textbf{Coordinates:} latitude and longitude used to identify the precise crash location. 2) \textbf{Crash:}  type, environmental conditions, road topology, and sequence of events that led to this crash. 3) \textbf{Vehicle:} velocity, model, impact point and damage points, and hints about the vehicle trajectories (e.g., ``going straight,'' ``turning left'').  

\noindent \textbf{Validation.}
 To improve the reconstruction quality, our implementation enforces a completeness constraint on the report, verifying the presence of geospatial coordinates of the crash location, road topology, and the category of pre-crash trajectory (as shown in Table \ref{tab:crash_coverage}). Reports that fail to meet this constraint are discarded since their reproduction cannot be adequately approximated. Further filtering constraints can be easily added to this stage. For example, to facilitate the curation of our benchmark of scenarios, we further constrain the pipeline to only keep dual-vehicle crashes. 
\vspace{-3pt}
\subsection{Map Reconstruction}
For the simulation environment to mirror the real-world location, the reconstruction process uses the \textbf{Retriever} module to query the free open map database OpenStreetMap (OSM)\cite{osm} to obtain  the relevant road network  for the crash site. The module prunes the retrieved data, using the OSMIUM map manipulation library \cite{osmium}, to reduce the map footprint  to facilitate its  upload into CARLA.  

Next, the \textbf{Converter} module transforms this topological OSM data into the OpenDRIVE format required by the CARLA simulator. 
This requires two transformations. 
First, convert the map's geospatial coordinates into an intermediate format that accounts for CARLA's flat world encoding.

Second,  convert that intermediate representation into the CARLA format using the CARLA built-in \textit{osm2odr()} function\cite{carla_docs}. Using this function is convenient but it has a cost: it flattens road networks, disregarding   vertical geometry. Consequently, our pipeline does not currently support any overpasses or tunnels, as addressing this limitation would require developing another custom conversion engine to replace the native tool. We leave the inclusion of maps with vertical variations for future work.

To support the simulation of crashes, the maps need one final transformation to overcome CARLA restrictions associated with certain road topologies. 
CARLA treats a road as a series of individual segments and it constrains high-risk maneuvers on certain segments which would prevent us from reproducing some crashes. For example, CARLA prohibits a vehicle from moving across solid lanes or driving against traffic, which are necessary to reproduce front-to-front collisions and sideswipes. This transformation  unifies disjointed lane segments into single, multi-lane road objects that enable lateral transitions and wrong-way driving. Making this kind of change provides the flexibility to faithfully reconstruct the specific vehicle behaviors observed in the crash reports.

\noindent \textbf{Validation.} Given the complexity of the OSM maps and the necessary transformations, and the possibility that the reports include inconsistent data, after converting to OpenDRIVE maps, \tool checks the accuracy of the resulting map. This validation selects 5 locations distributed on the map and converts them into CARLA coordinates. If the relative distance between those coordinates is the same in the original OSM map and the transformed one for CARLA, then the road geometry was maintained; otherwise, \tool highlights the error and excludes the report. We also discard those maps where the crash point is outside the roads, as those are either misreported or correspond to other types of crashes that we are not able to simulate in CARLA yet (i.e., in a parking lot). 

\subsection{Scenario  Reconstruction}

A critical challenge in using crash reports is the lack of precise data about the state of vehicles during the trajectory leading to the crash. Reproducing the crashes requires for such states to be estimated based on partial and often biased descriptions of event sequences and more objective but still partial data regarding the final state of the vehicles and their collision points. 

To perform such estimation, the \textit{Scene Constructor} relies on the \textbf{State Estimator}. This module consumes the output of the \textit{Extractor} and the \textit{map layout} from the \textit{Converter}. 
It then processes these data to identify the specific road segments and junction boundaries (for intersection crashes) that encompass the set of potential initial vehicle positions.
Then, the module invokes an off-the-shelf Large Language Model (LLM) (we used gemini-3-flash-preview for the generation of the benchmark) to infer the vehicle's initial position from the set. Note that \tool's architecture is LLM-agnostic, allowing users to swap models as long as they support structured reasoning. The insight of using an LLM is that large models trained with enormous datasets have encoded the capacity to fill the gaps in event sequences, rendering the likely missing initial states. The invocation prompt includes directives to derive initial vehicle position by calculating backward trajectories derived from the reported collision speeds, angles and sequences of events. Additional directives are included to confine initial vehicle positions within the provided valid road geometry and enforce right-hand traffic regulations for lane assignment. The specific prompt, including the reasoning process and the output format, are available in the repository.
 
The module also incorporates a series of analytical checks whose results serve to provide a feedback loop to the LLM's estimation, checking that the vehicles' initial positions fall within road boundaries and that the vehicles' orientations are aligned with the ones in the report. In the event of an invalid estimation, the module re-prompts the LLM with the refined directive, including the error, to re-estimate, subject to a maximum number of retries. Upon successful validation, the module outputs a structured JSON object containing the precise Cartesian coordinates for the crash location and the initial positions  of both vehicles, mapped to their respective road IDs from the OpenDrive map. 

Once the state estimation is completed, the \textbf{Launcher}   transforms the estimated states into executable vehicle trajectories (sequence of waypoints in CARLA WayPoint class format). To reconstruct the vehicles' trajectories leading to the crash, the module takes the estimated initial  position of the vehicles as  spawning targets and the crash location as the destination for each vehicle. To generate the waypoints that define a vehicle's trajectory, instead of using CARLA's \textit{GlobalRoutePlanner}, which relies on the map's graph topology and thus cannot generate paths for ``illegal'' maneuvers like wrong way driving, the \textit{Launcher} implements an alternative method that generates waypoints based on the particular crash, spawns the vehicles with  the reported velocity and allows for crashes. 
The module then spawns the vehicles within the generated CARLA map and controls their motion to recreate the sequence of events described in the crash report. It also saves the vehicles' position, orientation, speed, and generated trajectory in a structured JSON format for later replaying the scene at will.

\noindent \textbf{Validation.} After the \textit{Launcher} completes the simulation of the vehicles, it records the output of the vehicles' crash sensor, the impact point, and the trajectory of the vehicles. Then, it validates the scenario reconstruction similarity to the real-world crash  based on three criteria: 1) simulated crash must occur within 5 meters of the reported location, 2) simulated impact points of each vehicle match the impact points mentioned in the reports with a permissible deviation of $\pm 2$ clock positions (12 represents the front of the vehicle and 6 the rear), and 3) simulated trajectories match  one of the three directions mentioned in the report (``turning left'', ``turning right'', or ``going straight''). 
These thresholds account for undocumented maneuvers, such as last-second evasive maneuvers, that shift vehicles' terminal headings and impact angles, making exact replication difficult.

\section{Benchmark}
\label{sec:benchmark}

\begin{figure} 
    \centering
    \begin{minipage}{.9\linewidth}
        \centering
        \begin{subfigure}{0.4\linewidth} 
            \centering
            \includegraphics[width=\linewidth]{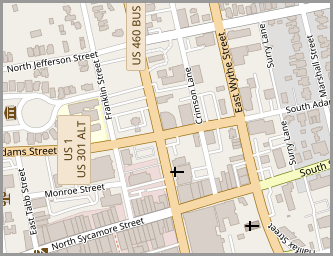} 
            \caption{Real-World}
            \label{fig:crash2_a} 
        \end{subfigure}
        \hfill
        \begin{subfigure}{0.4\linewidth}
            \centering
            \includegraphics[width=\linewidth]{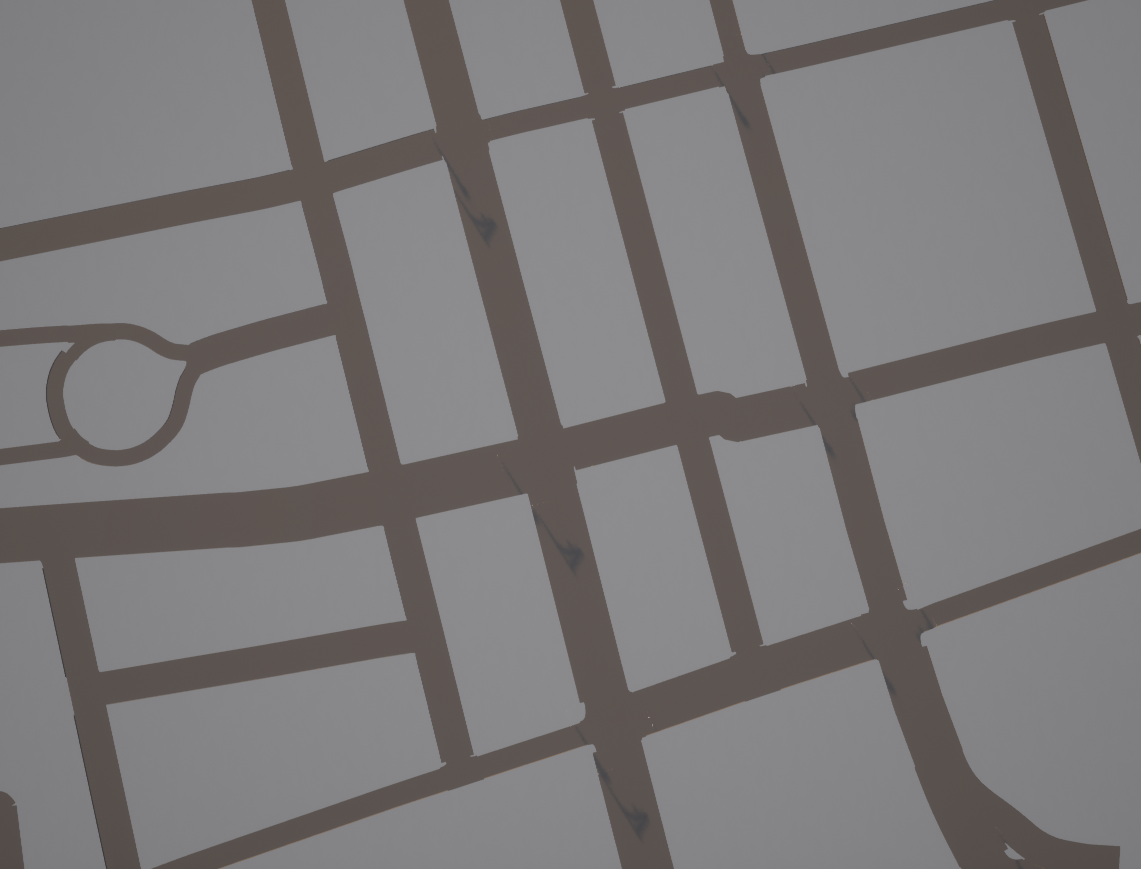}
            \caption{Carla-World}
            \label{fig:table}
        \end{subfigure}    
    \end{minipage}
    \caption{Road Topology of StateCase=510179 in (a) real-world map, (b) CARLA map reproduced by \tool } 
    \label{fig:toposim}
\end{figure}

Our second contribution   is a benchmark of crash scenarios built with \tool  and further curated to evaluate AVs. A key feature of the benchmark is the fidelity of the network  topology and signage where the vehicles' trajectories are situated. For every generated scenario, the AVs navigate roads that faithfully   represent the complexity of the accident site, avoiding the  oversimplifications of synthetic tracks. 
As an illustration, consider the crash report \footnote{\url{https://crashviewer.nhtsa.dot.gov/crashviewer/CrashAPI/crashes/GetCaseDetails?stateCase=510179\&caseYear=2023\&state=51\&format=xml}} stateCase=510179, which occurred at (Latitude: 37.22810833, Longitude: -77.40179167), in Petersburg, Virginia. \autoref{fig:toposim} shows the OpenStreetMap \cite{osm}  and the corresponding one generated through our pipeline and included in the benchmark.
The map topology, accurately converted from OpenStreetMap, shows the real-world lane widths and curvatures leading up to the site.

We started with an initial batch of 100 randomly sampled crash reports from NHTSA CrashView Database \cite{NHTSA_Crash_API}. \tool filtered 35 during the Map Reconstruction phase. The primary cause for exclusion was unsupported vertical geometry ($17$), followed by crash data issues such as `Incomplete Info' ($10$) and `Inconsistent crash location' ($8$). Of the 65 scenarios that successfully generated maps, $13$ failed to reproduce a collision during the simulation phase. This resulted in 52 verified crash scenarios. 
\begin{table} 
    \centering
    \small
    \begin{tabular}{lc}
        \toprule
        \textbf{Category} & \textbf{Count} \\
        \midrule
        \multicolumn{2}{l}{\textbf{Type of Collision}} \\
        \hspace{1em} Angle & 19 \\
        \hspace{1em} Front-to-Front & 23 \\
        \hspace{1em} Front-to-Rear & 3 \\
        \hspace{1em} Sideswipe, Opposite Direction & 2 \\
        \hspace{1em} Sideswipe, Same Direction & 0 \\
        \hspace{1em} Rear-to-Side & 0 \\
        \hspace{1em} Rear-to-Rear & 0 \\
        \hspace{1em} Others & 5 \\
        \midrule
        \multicolumn{2}{l}{\textbf{Road Topology}} \\
        \hspace{1em} Not an Intersection & 36 \\
        \hspace{1em} T-Intersection & 8 \\
        \hspace{1em} Four-way Intersection & 8 \\
        \hspace{1em} Y-Intersection & 0 \\
        \hspace{1em} Traffic Circle /  Roundabout & 0 \\
        \hspace{1em} Five-Point, or More & 0 \\
        \hspace{1em} L-Intersection & 0 \\
        \hspace{1em} Others & 0 \\
        \midrule
        \multicolumn{2}{l}{\textbf{Vehicle Trajectory}} \\
        \hspace{1em} Same Trafficway, Same Direction & 4 \\
        \hspace{1em} Same Trafficway, Opposite Direction & 27 \\
        \hspace{1em} Changing Trafficway, Vehicle Turning & 9 \\
        \hspace{1em} Intersecting Paths & 7 \\
        \hspace{1em} Others & 5 \\
        \bottomrule
    \end{tabular}
    \vspace{2ex}
    \caption{Crash Scenario Coverage for 52 reports in terms of Collision Type, Road Topology,  and Vehicle Trajectory}
          \vspace{-0.3in}
    \label{tab:crash_coverage}
\end{table}

We claim that our benchmark's strength lies in its  coverage of  crash scenarios along three dimensions: type of collisions, road topology, and trajectories. We validate this claim through the analysis of the 52 real-world accident reports reproduced as per their mapping to the categories defined for these dimensions in the Fatality Analysis Reporting System (FARS) Manual \cite{NHTSA_FARS_2025}. 
\autoref{tab:crash_coverage} provides the result of this mapping.  The benchmark covers 5 out of 8 collision types, with a prevalent distribution of Front-to-Front (23 cases) and Angle impacts (19 cases), while also encompassing Front-to-Rear and Sideswipe scenarios. These collisions are instantiated across varied topologies; the majority occur on non-intersection segments (36 cases), such as straight roads or curves, complemented by a balanced set of complex intersection scenarios, split evenly between T-Intersections (8) and Four-way Intersections (8). The analysis of vehicle trajectories reveals significant diversity in pre-crash dynamics, ranging from vehicles traveling in opposite directions (27 cases) to those involved in turning maneuvers (9) or intersecting paths (7). All uncovered categories  in Table \ref{tab:crash_coverage}, such as `Sideswipe, Same Direction', `Rear-to-Rear', and `Y-Intersection' were not present within the batch of 100 crash reports sampled for this analysis.

\section{Conclusion}
\label{sec:conclusion}

We introduce \tool, a pipeline that operationalizes real-world accident reports and road network data into high-fidelity CARLA scenarios. By automating the transition from accident databases to simulation, \tool ensures that generated scenarios retain the critical characteristics of crashes, moving beyond the limitations of synthetic benchmarks. 
Future work will expand \tool's support for complex topologies, such as tunnels and overpasses, collisions involving more than two participants, and pedestrians and cyclists.

\begin{acks}
This research is supported in part by NSF awards 2312487 and 2403060.
\end{acks}

\clearpage
\bibliographystyle{acm}
\bibliography{biblio}

\end{document}